\documentclass[jmp,preprint,showpacs,floatfix]{revtex4}
\usepackage{amsmath}
\usepackage{amsfonts}
\usepackage{amssymb}
\usepackage{dcolumn}
\usepackage{bm}
\usepackage{graphicx}
\newcommand{\graphr}{
\setlength{\unitlength}{1.8ex}
\begin{picture}(1.05,1)
\linethickness{1pt}
\put(.5,0){\circle*{.2}}
\put(0.5,.5){\circle{1}}
\end{picture}
}

\newcommand{\graphgamma}{
\setlength{\unitlength}{1.8ex}
\begin{picture}(1,1)(0,-0.25)
\linethickness{.5pt}
\put(0,0){\line(1,0){1}}
\put(0,0){\circle*{.2}}
\put(1,0){\circle*{.2}}
\end{picture}
}

\newcommand{\graphrra}{
\setlength{\unitlength}{1.8ex}
\begin{picture}(2,.5)(-.5,-0.25)
\put(0.5,0){\circle*{.2}}
\linethickness{.5pt}
\put(0,0){\circle{1}}
\put(1,0){\circle{1}}
\end{picture}
}

\newcommand{\graphrrb}{
\setlength{\unitlength}{1.8ex}
\begin{picture}(3,.5)(-0.25,.1)
\linethickness{1pt}
\put(0.50,0.00){\circle*{.2}}
\put(0.50,0.50){\circle{1}}

\put(2,0.00){\circle*{.2}}
\put(2,0.50){\circle{1}}

\end{picture}
}

\newcommand{\graphgra}{
\setlength{\unitlength}{1.8ex}
\begin{picture}(2.5,.5)(-0.25,-0.25)
\put(1,0){\circle*{.2}}
\put(2,0){\circle*{.2}}
\linethickness{1pt}
\put(0.5,0){\circle{1}}
\linethickness{.5pt}
\put(1,0){\line(1,0){1}}
\end{picture}
}

\newcommand{\graphgrb}{
\setlength{\unitlength}{1.8ex}
\begin{picture}(3.5,.5)(-0.25,-0.25)
\linethickness{1pt}
\put(1,0){\circle*{.2}}
\put(2,0){\circle*{.2}}
\put(3,0){\circle*{.2}}
\put(0.5,0){\circle{1}}

\linethickness{.5pt}
\put(2,0){\line(1,0){1}}
\end{picture}
}

\newcommand{\graphgga}{
\setlength{\unitlength}{1.8ex}
\begin{picture}(1.5,.5)(-0.25,-0.25)
\linethickness{.5pt}
\put(0,0){\circle*{.2}}
\put(1,0){\circle*{.2}}
\linethickness{1pt}
\put(0.5,0){\circle{1}}
\end{picture}
}

\newcommand{\graphggb}{
\setlength{\unitlength}{1.8ex}
\begin{picture}(1.5,.5)(-0.25,-0.25)
\linethickness{.5pt}
\put(0,-0.5){\circle*{.2}}
\put(0,-0.5){\line(1,0){1}}
\put(1,-.5){\circle*{.2}}
\put(1,-0.5){\line(0,1){1}}
\put(1,0.5){\circle*{.2}}
\end{picture}
}

\newcommand{\graphggc}{
\setlength{\unitlength}{1.8ex}
\begin{picture}(1.5,.5)(-0.25,-0.25)
\linethickness{.5pt}
\put(0,0.5){\line(1,0){1}}
\put(0,-0.5){\line(1,0){1}}
\linethickness{.5pt}
\put(0,0.5){\circle*{0.2}}
\put(0,-0.5){\circle*{0.2}}
\put(1,0.5){\circle*{0.2}}
\put(1,-0.5){\circle*{0.2}}
\end{picture}
}

\newcommand{\graphrrra}{
\setlength{\unitlength}{1.5ex}
\begin{picture}(3,1.5)(-1.5,-0.5)
\linethickness{0.001em}
\put(0,0){\circle*{0.5}}
\qbezier(0.00,0.00)(0.281,1.875)(1.125,0.750)
\qbezier(1.125,0.750)(1.969,-0.375)(0.00,0.00)
\qbezier(0.00,0.00)(-1.406,-1.125)(0.00,-1.125)
\qbezier(0.00,-1.125)(1.406,-1.125)(0.00,0.00)
\qbezier(0.00,0.00)(-1.969,-0.375)(-1.125,0.750)
\qbezier(-1.125,0.750)(-0.281,1.875)(0.00,0.00)
\end{picture}}

\newcommand{\graphrrrb}{
\setlength{\unitlength}{1.8ex}
\begin{picture}(3,1)(-0.25,-0.25)
\linethickness{1pt}
\put(0.5,0){\circle*{.2}}
\linethickness{1pt}
\put(2,0){\circle*{.2}}
\put(0.5,-0.5){\circle{1}}
\put(0.5,0.5){\circle{1}}
\put(2,0.5){\circle{1}}
\end{picture}
}

\newcommand{\graphrrrc}{
\setlength{\unitlength}{1.8ex}
\begin{picture}(4.50,1)(-0.25,0)
\linethickness{1pt}
\put(0.50,0.00){\circle*{.2}}
\put(0.50,0.50){\circle{1}}

\put(2,0.00){\circle*{.2}}
\put(2,0.50){\circle{1}}

\put(3.50,0.00){\circle*{.2}}
\put(3.50,0.50){\circle{1}}
\end{picture}
}

\newcommand{\graphgrra}{
\setlength{\unitlength}{1.8ex}
\begin{picture}(2,1)(-0.25,-0.25)
\linethickness{1pt}
\put(0.5,0){\circle*{.2}}
\put(1.5,0){\circle*{.2}}
\put(0.5,-0.5){\circle{1}}
\put(0.5,0.5){\circle{1}}

\linethickness{0.02em}
\put(0.5,0){\line(1,0){1}}
\end{picture}
}

\newcommand{\graphgrrb}{
\setlength{\unitlength}{1.8ex}
\begin{picture}(3.5,.75)(-0.25,-0.25)
\linethickness{1pt}
\put(1,0){\circle*{.2}}
\put(2,0){\circle*{.2}}
\put(0.5,0){\circle{1}}
\put(2.5,0){\circle{1}}

\linethickness{.5pt}
\put(1,0){\line(1,0){1}}
\end{picture}
}

\newcommand{\graphgrrc}{
\setlength{\unitlength}{1.8ex}
\begin{picture}(4,1)(-0.25,-0.25)
\linethickness{1pt}
\put(1,0){\circle*{.2}}
\put(2,0){\circle*{.2}}
\put(3,0){\circle*{.2}}
\put(0.5,0){\circle{1}}
\put(3,0.5){\circle{1}}

\linethickness{.5pt}
\put(1,0){\line(1,0){1}}
\end{picture}
}

\newcommand{\graphgrrd}{
\setlength{\unitlength}{1.8ex}
\begin{picture}(2.5,1)(-0.25,-0.25)
\linethickness{1pt}
\put(1,0.5){\circle*{.2}}
\put(0.5,-0.5){\circle*{.2}}
\put(1.5,-0.5){\circle*{.2}}
\put(0.5,0.5){\circle{1}}
\put(1.5,0.5){\circle{1}}

\linethickness{.5pt}
\put(0.5,-0.5){\line(1,0){1}}
\end{picture}
}

\newcommand{\graphgrre}{
\setlength{\unitlength}{1.8ex}
\begin{picture}(3,1)(-0.25,-0.25)
\linethickness{1pt}
\put(0.5,0){\circle*{.2}}
\put(2,0){\circle*{.2}}
\put(0.75,-0.5){\circle*{.2}}
\put(1.75,-0.5){\circle*{.2}}
\put(0.5,0.5){\circle{1}}
\put(2,0.5){\circle{1}}

\linethickness{.5pt}
\put(0.75,-0.5){\line(1,0){1}}
\end{picture}
}

\newcommand{\graphggra}{
\setlength{\unitlength}{1.8ex}
\begin{picture}(2.5,.75)(-0.25,-0.25)
\linethickness{1pt}
\put(0,0){\circle*{.2}}
\put(1,0){\circle*{.2}}
\put(0.5,0){\circle{1}}
\put(1.5,0){\circle{1}}
\end{picture}
}

\newcommand{\graphggrb}{
\setlength{\unitlength}{1.8ex}
\begin{picture}(3,.75)(-0.25,-0.25)
\linethickness{1pt}
\put(0,0){\circle*{.2}}
\put(1,0){\circle*{.2}}
\put(2,-0.5){\circle*{.2}}
\put(0.5,0){\circle{1}}
\put(2,0){\circle{1}}
\end{picture}
}

\newcommand{\graphggrc}{
\setlength{\unitlength}{1.8ex}
\begin{picture}(2.5,.75)(-0.25,-0.25)
\linethickness{.5pt}
\put(1,-0.25){\line(1,0){1}}
\put(1,-0.25){\line(-1,0){1}}

\put(1,-0.25){\circle*{.2}}
\put(0,-0.25){\circle*{.2}}
\put(2,-0.25){\circle*{.2}}
\linethickness{1pt}
\put(1,0.25){\circle{1}}
\end{picture}
}

\newcommand{\graphggrd}{
\setlength{\unitlength}{1.8ex}
\begin{picture}(2.5,.75)(-0.25,-0.25)
\linethickness{.5pt}
\put(0,0.75){\circle*{.2}}
\put(0,0.75){\line(0,-1){1}}
\put(0,-0.25){\circle*{.2}}
\put(0,-0.25){\line(1,0){1}}
\put(1,-0.25){\circle*{.2}}
\linethickness{1pt}
\put(1.5,-0.25){\circle{1}}
\end{picture}
}

\newcommand{\graphggre}{
\setlength{\unitlength}{1.8ex}
\begin{picture}(3,.75)(-0.25,-0.25)
\linethickness{.5pt}
\put(0,0.5){\circle*{.2}}
\put(0,0.5){\line(0,-1){1}}
\put(0,-0.5){\circle*{.2}}
\put(0,-0.5){\line(1,0){1}}
\put(1,-0.5){\circle*{.2}}
\put(2,-0.5){\circle*{.2}}
\linethickness{1pt}
\put(2,0){\circle{1}}
\end{picture}
}

\newcommand{\graphggrf}{
\setlength{\unitlength}{1.8ex}
\begin{picture}(2.5,1.25)(-0.25,-0.25)
\linethickness{.5pt}
\put(0,0.5){\line(1,0){1}}
\put(0,-0.5){\line(1,0){1}}
\linethickness{.5pt}
\put(0,0.5){\circle*{0.2}}
\put(0,-0.5){\circle*{0.2}}
\put(1,0.5){\circle*{0.2}}
\put(1,-0.5){\circle*{0.2}}
\put(1.5,0.5){\circle{1}}
\end{picture}
}

\newcommand{\graphggrg}{
\setlength{\unitlength}{1.8ex}
\begin{picture}(3,.75)(-0.25,-0.25)
\linethickness{.5pt}
\put(0,0.5){\line(1,0){1}}
\put(0,-0.5){\line(1,0){1}}
\linethickness{.5pt}
\put(0,0.5){\circle*{0.2}}
\put(0,-0.5){\circle*{0.2}}
\put(1,0.5){\circle*{0.2}}
\put(1,-0.5){\circle*{0.2}}
\put(1.5,0){\circle*{0.2}}
\put(2,0){\circle{1}}
\end{picture}
}

\newcommand{\graphggga}{
\setlength{\unitlength}{1.8ex}
\begin{picture}(1.5,1)(-0.25,-0.25)
\linethickness{.5pt}
\put(0,0){\circle*{.2}}
\put(0,0){\line(1,0){1}}
\put(1,0){\circle*{.2}}
\linethickness{1pt}
\put(0.5,0){\circle{1}}
\end{picture}
}

\newcommand{\graphgggb}{
\setlength{\unitlength}{1em}
\begin{picture}(1.5,.75)(-0.25,-0.25)
\linethickness{.5pt}
\put(0,-0.5){\circle*{.2}}
\put(0,-0.5){\line(1,1){1.0142}}
\put(0,-0.5){\line(1,0){1}}
\put(1,-.5){\circle*{.2}}
\put(1,-0.5){\line(0,1){1}}
\put(1,0.5){\circle*{.2}}
\end{picture}
}

\newcommand{\graphgggc}{
\setlength{\unitlength}{1.8ex}
\begin{picture}(2.5,.5)(-0.25,-0.25)
\linethickness{.5pt}
\put(0,0){\circle*{.2}}
\put(1,0){\circle*{.2}}
\put(1,0){\line(1,0){1}}
\put(2,0){\circle*{.2}}
\linethickness{1pt}
\put(0.5,0){\circle{1}}
\end{picture}
}

\newcommand{\graphgggd}{
\setlength{\unitlength}{1.8ex}
\begin{picture}(2.5,.75)(-0.25,-0.25)
\linethickness{.5pt}
\put(0,-0.5){\circle*{.2}}
\put(0,-0.5){\line(1,0){1}}
\put(1,-0.5){\circle*{.2}}
\put(1,-0.5){\line(1,0){1}}
\put(1,-0.5){\line(0,1){1}}
\put(1,0.5){\circle*{.2}}
\put(2,-0.5){\circle*{.2}}
\end{picture}
}

\newcommand{\graphggge}{
\setlength{\unitlength}{1.8ex}
\begin{picture}(1.5,1)(-0.25,-0.25)
\linethickness{.5pt}
\put(0,0.5){\circle*{.2}}
\put(0,0.5){\line(0,-1){1}}
\put(0,-0.5){\circle*{.2}}
\put(0,-0.5){\line(1,0){1}}
\put(1,-0.5){\circle*{.2}}
\put(1,-0.5){\line(0,1){1}}
\put(1,0.5){\circle*{.2}}
\end{picture}
}

\newcommand{\graphgggf}{
\setlength{\unitlength}{1.8ex}
\begin{picture}(1.5,1)(-0.25,-0.25)
\linethickness{.5pt}
\put(0,0.5){\circle*{.2}}
\put(0,-0.5){\circle*{.2}}
\put(1,0.5){\circle*{.2}}
\put(0,-0.5){\line(1,0){1}}
\put(1,-0.5){\circle*{.2}}
\linethickness{1pt}
\put(0.5,0.5){\circle{1}}\end{picture}
}

\newcommand{\graphgggg}{
\setlength{\unitlength}{1.8ex}
\begin{picture}(1.5,1)(-0.25,-0.25)
\linethickness{.5pt}
\put(0,0.75){\circle*{.2}}
\put(0,0.75){\line(1,0){1}}
\put(1,0.75){\circle*{.2}}
\put(1,0.75){\line(0,-1){1}}
\put(1,-0.25){\circle*{.2}}

\put(0,-0.75){\circle*{.2}}
\put(0,-0.75){\line(1,0){1}}
\put(1,-0.75){\circle*{.2}}
\end{picture}
}

\newcommand{\graphgggh}{
\setlength{\unitlength}{1.8ex}
\begin{picture}(1.5,1)(-0.25,-0.25)
\linethickness{.5pt}
\put(0,0.5){\circle*{.2}}
\put(0,0.5){\line(1,0){1}}
\put(1,0.5){\circle*{.2}}

\put(0,0){\circle*{.2}}
\put(0,0){\line(1,0){1}}
\put(1,0){\circle*{.2}}

\put(0,-0.5){\circle*{.2}}
\put(0,-0.5){\line(1,0){1}}
\put(1,-0.5){\circle*{.2}}
\end{picture}
}

\newcommand{\labeledgraphr}[1]{
\setlength{\unitlength}{1.8ex}
\begin{picture}(2,.75)(0,0.25)
\linethickness{1pt}
\put(1,.5){\circle*{.2}}
\put(1.5,0.25){\scriptsize{$#1$}}
\linethickness{1pt}
\put(0.5,0.5){\circle{1}}
\end{picture}
}

\newcommand{\labeledgraphgamma}[2]{
\setlength{\unitlength}{1.8ex}
\begin{picture}(2,.75)(-0.2,0)
\linethickness{1pt}
\put(0.25,0){\circle*{.2}}
\put(0.25,0){${}^{#1}$}
\put(1.25,0){\circle*{.2}}
\put(1.25,0){${}^{#2}$}
\linethickness{.5pt}
\put(0.25,0){\line(1,0){1}}
\end{picture}
}

\newcommand{\labeledgraphgra}[2]{
\setlength{\unitlength}{1.8ex}
\begin{picture}(2.5,1)(0,0)
\linethickness{1pt}
\put(0.5,1){\circle*{.2}}
\put(-0.5,1.5){\scriptsize $#1$}
\put(1.5,1){\circle*{.2}}
\put(2,0.25){\scriptsize $#2$}
\put(0.5,0.5){\circle{1}}
\linethickness{.5pt}
\put(0.5,1){\line(1,0){1}}
\end{picture}
}

\newtheorem{definition}{Definition}

\newtheorem{lemma}{Lemma}

\newcommand{\stackleft}[2]{ \,{}^{(#1)}_{\ #2}}
\newcommand{\stacklr }[5]{ \,{}^{(#1)}_{\ #2}#3^{#4}_{#5}\,}

\allowdisplaybreaks
\makeindex
\begin{document}
\title{A Complete Basis for a Perturbation Expansion of the General $N$-Body Problem}
\author{W. Blake Laing\footnote{Current Address: Department of Physics, Kansas State University}, David W. Kelle\footnote{Current Address: Department of Mathematics, Florida State University}, Martin Dunn, and Deborah K. Watson}
\affiliation{Homer L. Dodge Department of Physics and Astronomy, University of Oklahoma}
\date{\today}

\begin{abstract}
We discuss a basis set developed to calculate perturbation coefficients in an
expansion of the general $N$-body problem.  This basis has two advantages. 
First, the basis is complete
order-by-order for the perturbation series.  Second, the number of
independent basis tensors spanning the space for a given
order does not scale
with $N$, the number of particles, despite the generality of the problem. At
first order, the number of basis tensors is 23 for all $N$
although the problem at first
order scales as $N^6$. The perturbation series is expanded in inverse
powers of the spatial dimension. This results in a maximally symmetric
configuration at lowest order which has a point group isomorphic
with the symmetric group, $S_N$. The resulting perturbation series is
order-by-order invariant under the $N!$ operations of the $S_N$ point
group which
is responsible for the slower than exponential growth of the basis.
In this paper, we perform the first test of this formalism including the
completeness of the basis through first order by comparing to an exactly
solvable fully-interacting problem of $N$ particles with a two-body harmonic
interaction potential.
\end{abstract}

\maketitle

\section{Introduction}
In a previous paper\cite{wavefunction1}, we described the development of a perturbation method
for the general $S$-wave $N$-body problem through first order.  Group 
theoretic and
graphical techniques were used to describe the interacting
$N$-body wave function for a
system of identical bosons with general interactions.  Solutions for this
problem are known to scale exponentially
with $N$ requiring that resources be essentially doubled
for each particle added\cite{liu:2007,montina2008}.
As $N$ increases beyond a few tens of particles, this growth
makes a direct numerical
simulation intractable without approximations given current numerical
resources.  Typical approximations truncate the Hilbert space of the exact
solution by using a basis that spans a smaller Hilbert space or by
truncating a perturbation series or both\cite{MKR,Cederbaum1,Cederbaum2,Landau&Binder,holzmann:99,nilsen:05,blume:01,MSTTV,Nunes,banerjee:01,GBSS}.
With bosonic systems, various Monte Carlo approaches may be employed which scale in a polynomial fashion
with $N$\,, making larger-$N$ calculations feasible.\cite{Landau&Binder,holzmann:99,nilsen:05,MSTTV}

In Paper~\onlinecite{wavefunction1}, a perturbation series is developed in inverse powers of the
spatial dimension. This results in a maximally symmetric configuration at lowest order having a
point group isomorphic to the symmetric group $S_N$\,. The basis used is complete, at each order finite, and,
in fact small, having only twenty three members at first order, despite the $N^6$ scaling of the problem at this order.

In this paper, we test this formalism which truncates the perturbation series,
but determines each term in the series exactly using group theory and
graphical techniques. This perturbation series is order-by-order invariant under the
operations of the $S_N$ point group.  The tensor blocks needed at each order
can thus be decomposed into a basis also invariant under the $S_N$
point group.  It is
this restriction, i.e. the invariance under $N!$ permutations, that
slows the growth of the number of basis tensors
as $N$ increases, resulting in
a basis that is small in contrast to the $N^6$ growth of the vector space
at first order.

We have named the basis tensors, ``binary invariants.''  ``Binary'' because
the elements within a basis tensor are ones or zeros; ``invariants'' because
the basis tensors are invariant under $N!$ permutations of the particles
labels.

This strategy effectively separates the $N$ scaling away from the rest of
the physics and then tackles the $N$-scaling problem using the symmetry
of the $S_N$ group. The full problem, of course, scales exponentially
in $N$, so as higher orders of the perturbation series are included the
full exponential $N$ scaling of the problem will appear.
However, in this methodology,
the $N$ scaling problem has been compartmentalized away from the rest of the
physics and dealt with using group theoretic and graphical techniques; i.e.
it becomes a straight mathematical problem.
Once this mathematical work, which involves significant analytical effort,
has been completed at a given order, it never has to be repeated
again for a new interaction or a different number of particles,

The formalism being tested in this paper has been presented in a series of
papers beginning with
the isotropic, lowest-order ground-state
wavefunction~\cite{normalmode},
the isotropic, lowest-order ground-state density profile~\cite{density0},
and the isotropic, first-order ground-state wavefunction~\cite{wavefunction1}.

In Section~\ref{sec:PGS} we review the large-dimension point group symmetry and discuss its implications in perturbation theory.
In Section~\ref{app:Bin} we discuss the binary invariants themselves and the graphs which label them, and in Section~\ref{sec:WFTFO}
we very briefly review the general theory for the wave function through first order from Paper~\onlinecite{wavefunction1}. In Section~\ref{sec:Test}
we compare the results of the general theory for the wave function through first order applied to the problem of the harmonically
interacting system under harmonic confinement with an expansion of the exact interacting wave function through first order.
This work is performed in Appendix~\ref{sec:HarmIntWF} where we exactly solve for the wave function of the harmonically interacting $N$-particle problem
under harmonic confinement in $D$ dimensions. This exact solution is then expanded through first order in the square root of the
inverse dimension of space.

\section{The Large-Dimension, Point Group Symmetry and it's Implications In Perturbation Theory}
\label{sec:PGS}
As discussed in previous papers on dimensional perturbation theory (DPT) the Hamiltonian and
Jacobian-weighted wave function and energy are expanded in powers of $\delta^{1/2}$\,:
\begin{equation} \label{eq:dpt_exp}
\renewcommand{\arraystretch}{2} \begin{array}{r@{}l@{}c}
{\displaystyle \bar{H} = \bar{H}_{\infty} + \delta^{\frac{1}{2}}
\, \bar{H}_{-1} + \delta} & {\displaystyle \, \sum_{j=0}^\infty
\left(\delta^{\frac{1}{2}}\right)^j 
\bar{H}_j } & \\
{\displaystyle \Phi(\bar{r}_i,\gamma_{ij}) = } & {\displaystyle \,
\sum_{j=0}^\infty \left(\delta^{\frac{1}{2}}\right)^j 
\Phi_j }& \\
{\displaystyle \bar{E} = \bar{E}_{\infty} + \delta^{\frac{1}{2}}
\, \bar{E}_{-1} + \delta} & {\displaystyle \, \sum_{j=0}^\infty
\left(\delta^{\frac{1}{2}}\right)^j 
\bar{E}_j } &
 \,,
\end{array}
\renewcommand{\arraystretch}{1}
\end{equation}
where
\begin{eqnarray}
 \bar{H}_{\infty} & = & \bar{E}_{\infty}  \\
\bar{H}_{-1} & = &  \bar{E}_{2n-1} =  0\,, \label{eq:mone_H} \\
\bar{H}_{0} & = & -\frac{1}{2}\stacklr{0}{2}{G}{}{\nu_1,\nu_2}
\partial_{{\bar{y}^\prime}_{\nu_1}}
\partial_{{\bar{y}^\prime}_{\nu_2 }} +
\frac{1}{2} 
\stacklr{0}{2}{F}{}{\nu_1,\nu_2}
\bar{y}^\prime_{\nu_1 } \label{eq:harm_H}
\bar{y}^\prime_{\nu_2}+\stacklr{0}{0}{F}{}{} \,, \label{eq:H0y} \\
\bar{H}_{1} & = &-\frac{1}{2}\stacklr{1}{3}{G}{}{\nu_1,\nu_2,\nu_3}
\bar{y}^\prime_{\nu_1 } \partial_{{\bar{y}^\prime}_{\nu_2 }}
\partial_{{\bar{y}^\prime}_{\nu_3 }}
-\frac{1}{2}\stacklr{1}{1}{G}{}{\nu}
\partial_{{\bar{y}^\prime}_\nu}
+\frac{1}{3!}\stacklr{1}{3}{F}{}{\nu_1,\nu_2,\nu_3}
\bar{y}^\prime_{\nu_1 } \bar{y}^\prime_{\nu_2}
\bar{y}^\prime_{\nu_3 } +\stacklr{1}{1}{F}{}{\nu}
\bar{y}^\prime_\nu \,. \label{eq:one_H}
\end{eqnarray}
The internal displacement coordinate vector $\bar{y}^\prime_{\nu}$ is a column vector composed of the displacement radii $\bar{r}^\prime_{i}$ and the displacement angle cosines $\bar{\gamma}^\prime_{i,j}$:
\begin{equation}\label{eq:ytransposeP}
\begin{array}[t]{l} {\bar{\bm{y}}'} = \left( \begin{array}{c} r \\
\overline{\bm{\gamma}}' \end{array} \right) \,, \;\;\;
\mbox{where} \;\;\; \\ \mbox{and} \;\;\; \bar{\bm{r}}' = \left(
\begin{array}{c}
\bar{r}'_1 \\
\bar{r}'_2 \\
\vdots \\
\bar{r}'_N
\end{array}
\right) \,, \end{array} 
\overline{\bm{\gamma}}' = \left(
\begin{array}{c}
\overline{\gamma}'_{12} \\ \cline{1-1}
\overline{\gamma}'_{13} \\
\overline{\gamma}'_{23} \\ \cline{1-1}
\overline{\gamma}'_{14} \\
\overline{\gamma}'_{24} \\
\overline{\gamma}'_{34} \\ \cline{1-1}
\overline{\gamma}'_{15} \\
\overline{\gamma}'_{25} \\
\vdots \\
\overline{\gamma}'_{N-2,N} \\
\overline{\gamma}'_{N-1,N} \end{array} \right) \,,
\end{equation}
and is related to the dimensionally-scaled internal coordinates by
\begin{equation} \label{eq:taylor1}
\bar{r}_{i} = \bar{r}_{\infty} +\delta^{1/2}\,\bar{r}'_{i} \;\;\;
\mbox{and} \;\;\; \gamma_{ij} =
\overline{\gamma}_{\infty} +\delta^{1/2}\,\overline{\gamma}'_{ij} \,,
\end{equation}
where the $\bar{r}_{i}$ are the dimensionally scaled radii and the $\gamma_{ij}$ are the angle cosines between the position
vectors of the $N$ particles.

The superprescript on the $\bm{F}$ and $\bm{G}$ tensors in
parentheses in Eqs.~(\ref{eq:mone_H}), (\ref{eq:H0y}) and (\ref{eq:one_H}) denotes the order in $\delta^{1/2}$ that the term enters
(harmonic being zeroth order). The subprescripts denote the rank of
the tensors.

In general, $\bar{H}_{n}$ is of order $n+2$ in the elements and derivatives of ${\bar{\bm{y}}'}$ ($2^{\rm nd}$ order in the
derivatives), and formed of either all even or all odd powers of the elements and derivatives of ${\bar{\bm{y}}'}$ when $n$
is even or odd respectively.

According to Eq.~(\ref{eq:taylor1}) the system localizes as $D \rightarrow \infty$ on a configuration centered about
$\bar{r}_{i} = \bar{r}_{\infty}$ and $\gamma_{ij} = \overline{\gamma}_{\infty}$. This structure has the highest
degree of symmetry where all particles are equidistant
from the center of the trap and equiangular from each other (a
configuration that's only possible in higher dimensions). The point
group of this structure is isomorphic to $S_N$ which in effect interchanges the
particles in the large dimension structure. This together with the
fact that the full $D$-dimensional Hamiltonian, $\hat{H}$\,, is invariant under
particle exchange means the DPT expansion of
Eq.~(\ref{eq:dpt_exp}) is order-by-order invariant under this $S_N$
point group, i.e.\ the $\bar{H}_j$ are each invariant under the
$S_N$ point group. This greatly restricts the $\bm{F}$ and $\bm{G}$
tensors of Eqs.~(\ref{eq:harm_H}) and (\ref{eq:one_H}). In three dimensions a corresponding $N$-particle structure
would have a point group of lower symmetry, i.e.\ one not isomorphic to $S_N$ despite the fact that all
of the particles are identical. It is this profound restriction on the $\bm{F}$ and $\bm{G}$ tensors from the $S_N$ point group
symmetry, which in itself is a direct consequence of developing a perturbation theory about the large-dimension limit,
that allows for an essentially
analytic solution of the problem at a given order in the perturbation theory for  {\em any} $N$\,. For example, with a third-rank tensor,
instead of $\left(N(N+1)/2\right)^3$ possible independent elements (a number which becomes quickly intractable since, for example,
with only ten particles this would be more that a million elements), the $S_N$ point group symmetry restricts
the number of independent elements to only twenty three, independent of the value of $N$\,.

\subsection{The Reducibility of the $\bm{F}$ and $\bm{G}$ Tensors under $\bm{S}_{\bm{N}}$}
\label{subsec:RFGTS}
The maximally symmetric point group $S_N$\,, together with the invariance of the full Hamiltonian
under particle interchange, requires that the $F$
and $G$ tensors be invariant under the interchange of particle
labels (the $S_N$ group). In fact the various blocks of the $F$ and $G$
tensors are themselves invariant under particle interchange induced by the point group.
For example, $\stacklr{1}{3}{F}{rr\gamma}{i,j,(kl)}$ is never transformed
into $\stacklr{1}{3}{F}{\gamma\gamma\gamma}{(ij),(kl)(mn)}$\,, i.e.\ the $r$
and $\gamma$ labels are preserved since the $S_N$ group does not transform
an $\bar{r}'_i$ coordinate into a $\overline{\gamma}'_{(ij)}$ coordinate.

The various $r-\gamma$ blocks of the $F$ and $G$ tensors may be decomposed into invariant, and, this time irreducible, blocks. Thus for example, $\stacklr{0}{2}{Q}{}{}$ may be decomposed into the blocks
\begin{equation}
\begin{array}{ll}
\stacklr{0}{2}{Q}{rr}{i,i} &
\hspace{2ex} \forall \hspace{2ex} i \\
\stacklr{0}{2}{Q}{rr}{i,j} & 
\hspace{2ex} \forall \hspace{2ex} i \neq j  \\
\stacklr{0}{2}{Q}{r\gamma}{i,(ik)} & \hspace{2ex} \forall
\hspace{2ex} i < k\,,  \\
\stacklr{0}{2}{Q}{r\gamma}{i,(jk)} &  \hspace{2ex} \forall \hspace{2ex}
i \neq j < k \\
\stacklr{0}{2}{Q}{\gamma\gamma}{(ij),(ij)} & \hspace{2ex}
\forall \hspace{2ex} i < j  \\
\stacklr{0}{2}{Q}{\gamma\gamma}{(ij),(ik)} & \hspace{2ex}
\forall \hspace{2ex} i < j \neq k > i \\
\stacklr{0}{2}{Q}{\gamma\gamma}{(ij),(kl)} & \hspace{2ex} \forall
\hspace{2ex} \begin{array}[t]{l} i \neq j \neq k \neq l\,, \\ i <
j\,, \hfill k < l \end{array}
\end{array}
\end{equation}
all of which remain disjoint from one another under the $S_N$ group.
Significantly, invariance under particle interchange requires that tensor elements related by a label permutation induced 
by the point group must be equal. This requirement partitions the set of tensor elements for each block into disjoint
subsets of identical elements. Consider the elements of the $\stackleft{0}{2}Q^{rr}$ block. The element
$\stacklr{0}{2}{Q}{rr}{1,1}$ belongs to a set of $N$ elements (of the form $\stacklr{0}{2}{Q}{rr}{i,i}$) which are
related by a permutation induced by the point group, and therefore must have
equal values. Likewise, the element $\stacklr{0}{2}{Q}{rr}{1,2}$ belongs to a set of $N-1$ elements related by a
permutation induced by the point group and sharing a common value. Proceeding in this fashion, we
observe that the blocks of the lowest-order tensors are partitioned into the following set of identical
elements which remain disjoint under particle interchange:
\begin{eqnarray}\label{eq:Q0partition}
\stacklr{0}{2}{Q}{rr}{i,i} & = & \stacklr{0}{2}{Q}{rr}{k,k}
\hspace{2ex} \forall \hspace{2ex} i \mbox{ and } k \\
\stacklr{0}{2}{Q}{rr}{i,j} & = & \stacklr{0}{2}{Q}{rr}{k,l}
\hspace{2ex} \forall \hspace{2ex} i \neq j \mbox{ and } k \neq l \\
\stacklr{0}{2}{Q}{r\gamma}{i,(ik)} & = &
\stacklr{0}{2}{Q}{r\gamma}{l,(lm)} =
\stacklr{0}{2}{Q}{r\gamma}{n,(pn)} \hspace{2ex} \forall
\hspace{2ex} i < k\,, \hspace{2ex} l < m\,, \mbox{ and } n
> p \\
\stacklr{0}{2}{Q}{r\gamma}{i,(jk)} & = &
\stacklr{0}{2}{Q}{r\gamma}{l,(mn)} \hspace{2ex} \forall \hspace{2ex}
i \neq j < k \mbox{ and } l
\neq m < n \\
\stacklr{0}{2}{Q}{\gamma\gamma}{(ij),(ij)} & = &
\stacklr{0}{2}{Q}{\gamma\gamma}{(kl),(kl)} \hspace{2ex}
\forall \hspace{2ex} i < j \mbox{ and } k < l \\
\stacklr{0}{2}{Q}{\gamma\gamma}{(ij),(ik)} & = &
\stacklr{0}{2}{Q}{\gamma\gamma}{(lm),(ln)} \hspace{2ex}
\forall \hspace{2ex} i < j \neq k > i  \mbox{ and } l < m \neq n >l \\
\stacklr{0}{2}{Q}{\gamma\gamma}{(ij),(kl)} & = &
\stacklr{0}{2}{Q}{\gamma\gamma}{(mn),(pq)} \hspace{2ex} \forall
\hspace{2ex} \begin{array}[t]{l} i \neq j \neq k \neq l \\ i <
j\,, \hfill k < l \end{array} \mbox{ and }
\begin{array}[t]{l} m \neq n \neq p \neq q \\ m < n\,, \hfill p < q \end{array}
\label{eq:Q0partitiongg}
\end{eqnarray}
%
The lowest-order block matrices contain the sets of elements in Eqs.~(\ref{eq:Q0partition})-(\ref{eq:Q0partitiongg})
arranged in an intricate pattern. A similar, but more involved partitioning occurs for higher rank $\bm{F}$ and $\bm{G}$
tensors.

In Paper~\onlinecite{wavefunction1} it was shown that this decomposition could be expressed in terms of binary tensors which are invariant under $S_N$
and are labelled by graphs. We term these binary tensors, binary invariants.

\section{Binary Invariants}
\label{app:Bin}
\subsection{Introducing Graphs}
\label{sec:IntroGraph}
\begin{definition}
A \emph{graph} $\mathcal{G} =(V,E)$ is a set of vertices $V$ and
edges $E$. Each edge has one or two associated vertices, which are
called its \emph{endpoints}\cite{gross:04}.
\end{definition}

For example, 
\graphggrc is a graph $\mathcal{G}$ with three vertices and
three edges. We allow our graphs to include loops and multiple edges\cite{multigraph}.
A graph contains information regarding the connectivity of edges and vertices only: the
orientation of edges and vertices is insignificant. 

We introduce a mapping which associates each tensor element with a graph as
follows:
\begin{enumerate}
\item draw a labeled vertex ($\bm{\cdot}$ $i$
) for each distinct index in the set of indices of the element
\item draw an edge ($\labeledgraphgamma{i}{j}$) for each double index $(ij)$
\item draw a ``loop'' edge ($\labeledgraphr{i}$) for each distinct single index $i$
\end{enumerate} 
For example, the graph corresponding to the tensor element $\stacklr{0}{2}{Q}{r\gamma}{i,(ij)}$ under this mapping is $\labeledgraphgra{i}{j}$\,.

Two graphs with the same number of vertices and edges that are connected in the same way are called \emph{isomorphic}. The elements of the $S_N$ group are permutation operations which interchange particle labels. Two elements with graphs that are not isomorphic are never related by a permutation of $S_N$
group. We label each set of isomorphic graphs by a graph with no vertex labels. Denoting the set of unlabeled graphs for each block as $\mathbb{G}_{X_1X_2\ldots X_n}$, where $n$ is the rank of the tensor block (and therefore the number of edges in each graph in the set) and $X$ is $r$ or $\gamma$\,,
we have
\begin{eqnarray}\label{eq:GXX}
\mathbb{G}_{rr}&=&\{\graphrra,\graphrrb\}
\nonumber\\
\mathbb{G}_{\gamma r}&=&\{\graphgra,\graphgrb\}
\\
\mathbb{G}_{\gamma \gamma}
&=&\{\graphgga,\graphggb,\graphggc\}
\nonumber
\end{eqnarray}
\begin{eqnarray}\label{eq:GXXX}
\mathbb{G}_{r}&=&\{\graphr\}
\nonumber\\
\mathbb{G}_{\gamma}&=&\{\graphgamma\}
\nonumber\\
\mathbb{G}_{rrr}&=&\{\graphrrra,\graphrrrb,\graphrrrc\}
\\
\mathbb{G}_{\gamma rr}
&=&\{\graphgrra,\graphgrrb,\graphgrrc,\graphgrrd,\graphgrre\}
\nonumber\\
\mathbb{G}_{\gamma \gamma r}
&=&\{\graphggra,\graphggrb,\graphggrc,\graphggrd,\graphggre,\graphggrf,\graphggrg\}
\nonumber\\
\mathbb{G}_{\gamma \gamma \gamma}
&=&\{\graphggga,\graphgggb,\graphgggc,\graphgggd,\graphggge,\graphgggf,\graphgggg,\graphgggh\}
\nonumber
\end{eqnarray}

Now consider a tensor, for which all of the elements labeled by a a single isomorphic set of
graphs are equal to unity, while all of the other elements labeled by graphs heteromorphic to this single
set of isomorphic graphs are equal to zero. We term this tensor a binary invariant,
$[B ({\mathcal G})]_{\nu_1, \nu_2, \ldots}$ since it
is invariant under the $S_N$ group, and we label it by the graph $\mathcal{G}$\., sans particle labels
at the vertices, for the non-zero elements all of which are equal to unity. Thus each of the above graphs
denotes a binary invariant, $[B ({\mathcal G})]_{\nu_1, \nu_2, \ldots}$\,. Explicit expressions for these binary
invariants for arbitrary $N$ may be found in the EPAPS document~\cite{JMPEPAPS}.

\subsection{Binary invariants are a basis}
That the small number of binary invariants are a complete basis with which to represent any $S_N$ invariant tensor in
the $N(N+1)/2$-dimensional $\bar{\bm{r}}'$-$\bar{\bm{\gamma}}'$ space can be seen as follows.

\begin{lemma}
 The set of binary invariants, $[B ({\mathcal G})]_{\nu_1, \nu_2, \ldots}$\,, for all $\mathcal{G}\in\mathbb{G}$ are linearly independent
\end{lemma}
\begin{description}
\item[Proof]This follows from the fact that no binary invariant shares a non-zero element with another binary invariant for a different graph.
\end{description}
\begin{lemma}
 The set of binary invariants for all $\mathcal{G}\in\mathbb{G}$ spans the invariant tensor space.
\end{lemma}
\begin{description}
\item[Proof]From the mapping of Sec.~\ref{sec:IntroGraph} above, relating  graphs to tensor elements in the $\bar{\bm{r}}'$-$\bar{\bm{\gamma}}'$
space, every tensor element is related to an unlabeled graph, and so the binary invariants for all possible unlabeled graphs span the
invariant tensor subspace.
\end{description}
\newtheorem{theorem}{Theorem}
\begin{theorem}
 The set of binary invariants $\{ B\left(\mathcal{G}\right):\mathcal{G} \in \mathbb{G}\}$ forms a basis for the DPT Hamiltonian coefficient tensors.
\end{theorem}
\begin{description}
 \item[Proof]This result follows from the definition of a basis and that the set of binary invariants for all $\mathcal{G}\in\mathbb{G}$ is linearly independent and spans the vector space.
\end{description}

As shown in Reference~\cite{kellecapstone} this result generalizes to any group with any set of tensors invariant under that group.

Therefore, we may resolve the $Q$ tensors at any order as a finite linear combination of binary invariants:

\begin{equation}\label{eq:graphresolution}
\stacklr{O}{R}{Q}{block}{\nu_1,\nu_2,\ldots,\nu_R} = \sum_{\mathcal{G}\in
\mathbb{G}_{block}}Q^{block}(\mathcal{G}) \, \left[B^{block}(\mathcal{G})\right]_{\nu_1,\nu_2,\ldots,\nu_R}
\end{equation}
where $\mathbb{G}_{block}$ represents the set of graphs present in the order-$O$, rank-$R$ tensor block
$\stacklr{O}{R}{Q}{block}{}$, and the binary invariant $B^{block}(\mathcal{G})$ has the same dimensions as the original $Q$ tensor block. The scalar quantity $Q^{block}(\mathcal{G})$ is the expansion
coefficient. 

The resolution of symmetric tensor blocks in the basis of binary invariants in Eq.~(\ref{eq:graphresolution}) represents
a generalization of a technique used at lowest order in Refs.~\onlinecite{normalmode} and \onlinecite{GFpaper} to arbitrary order.
This equation also separates the specific interaction dynamics present in $Q^{block}(\mathcal{G})$ from
the point group symmetry embodied in $B^{block}(\mathcal{G})$\,.

\section{The Wave Function Through First Order}
\label{sec:WFTFO}
\subsection{Lowest-Order Ground State Wavefunction}
Since the Hamiltonian of the lowest-order wave function (Eq.~(\ref{eq:H0y})) has the form of a $N(N+1)/2$-dimensional coupled harmonic
oscillator, the lowest-order wave function will be a product of one-dimensional, harmonic-oscillator, normal-mode functions.

The lowest-order DPT wave function, ${}_g \hspace{-0.25em} \Phi_0 (\mathbf{q'})$, for the ground state is given by
 \begin{equation}
     \label{eq:groundWF}{}_g \hspace{-0.25em} \Phi_0 (\mathbf{q'}) = \prod_{\nu = 1}^{P}
     \hspace{0.25em} \hspace{0.25em} \phi_0 
     \left( \sqrt{\bar{\omega}_{\nu}} \,
     \hspace{0.25em} q'_{\nu} \right),
   \end{equation}
where
   \begin{equation}
     \label{eqphi0} \phi_0 
     \left(
     \sqrt{\bar{\omega}_{\nu}} \,q'_{\nu} \right) = \left(
     \frac{\bar{\omega}_{\nu}}{\pi} \right)^{\frac{1}{4}}\exp \left( - \frac{1}{2}  \bar{\omega}_{\nu}\,{q'_{\nu}}^2 \right)\,.
   \end{equation} 
There are $N(N+1)/2$ normal modes and up to $N(N+1)/2$ distinct frequencies, a number which would become impossibly large to solve
for if it weren't for the $S_N$ point group symmetry expressed in the invariance of the $\bm{F}$ and $\bm{G}$ tensors, and the small,
$N$-independent number of binary invariants spanning the invariant tensor spaces. In Refs.~\onlinecite{normalmode}, \onlinecite{GFpaper} and \onlinecite{energy} we have used this $S_N$
point group symmetry to derive both the frequencies and normal modes of the lowest-order, Jacobian-weighted wave function for
arbitrary $N$\,. This analysis results in only five distinct frequencies, associated with center-of-mass and breathing modes,
radial and angular singly-excited state modes, and phonon modes. Each of these frequencies is associated with a set of normal
modes which transforms under an irreducible representation of the $S_N$ point group.

\subsection{First-Order Wave Function}
Using the $S_N$ point group symmetry expressed in the invariance of the $\bm{F}$ and $\bm{G}$ tensors, and the small,
$N$-independent number of binary invariants spanning the invariant tensor spaces, in Paper~\onlinecite{wavefunction1} we have also derived
the first-order correction to the lowest-order, harmonic wavefunction. If we write
\begin{equation}\label{eq:Phi1hat}
 \Phi (\mathbf{q'}) = (1 + \delta^{\frac{1}{2}} \hat{\Delta})
   \Phi_0 (\mathbf{q'}) + O(\delta) \,,
\end{equation}
then $\hat{\Delta}$ satisfies the commutator eigenvalue equation
\begin{equation} \label{eq:deltahatcommute}
 [\hat{\Delta},\hat{H}_0]\Phi_0=\hat{H}_1\Phi_0.
\end{equation}
To solve this equation, we note that since $\Phi_0 (\mathbf{q'})$ is a Gaussian function, the
derivatives in $\hat{H}_1$ and $\hat{H}_0$ written in normal coordinates "bring down"
normal coordinates from the exponent so that
$\hat{H}_1$ effectively becomes a 3rd-order polynomial of only odd powers in $\mathbf{q'}$.
Then from Eq.~(\ref{eq:deltahatcommute}) $\hat{\Delta}$ is a cubic polynomial and of only odd powers in the normal modes.
When $\hat{\Delta}$ is re expressed in terms of internal displacement coordinates, $\bm{r}'$ and $\bm{\gamma}'$\,,
it is cubic and of only odd powers in these internal displacement coordinates.

The ground-state wave function is also scalar under $S_N$, and so when it is expressed in terms of internal displacement coordinates
it involves binary invariants which take powers of the internal displacement coordinates and couple them together to produce
a scalar under $S_N$\,.

\section{A Test of the Theory: The Harmonically Confined, Harmonically Interacting System}
\label{sec:Test}
The general theory developed in Paper~\onlinecite{wavefunction1} and and Ref.~\onlinecite{normalmode}, and briefly reviewed in this paper, is extensive, and we test it on a non-trivial, interacting,
analytic solvable model: the harmonically-interacting system of $N$ particles under harmonic confinement:
\begin{equation}
H = \frac{1}{2} \, \left( \sum_i^N  \left[ - \frac{\partial^2}{\partial \bm{r}_i^2} + \omega_t^2 \bm{r}_i^2 \right]  + \sum_{i<j}^{N} \omega_p^2 \bm{r}_{ij}^2 \right) \,.
\end{equation}

\subsection{The Wavefunction Through First Order}
In Appendix~\ref{sec:HarmIntWF} we solve the harmonically-confined, harmonically-interacting system of $N$ particles exactly for
the ground-state wave function (see  Eq.~(\ref{eq:WFNJ})), and then from this derive the exact analytic perturbation series for the $N$-body wavefunction (weighted by a Jacobian) through first order:
\begin{eqnarray} \label{eq:wf1storder}
\Psi_J &=& \left(\frac{1}{\sqrt[4]{\pi}} \right)^{\frac{N(N+1)}{2} } \left( 1 + \frac{1}{2}\delta^{\frac{1}{2}} \Delta_{\mathbf{\bar{y}}^\prime}+O(\delta)\right) \exp{ \left( -\left[\mathbf{\bar{y}}^\prime\right]^T\,\mathbf{\bar{\Omega }}_{\mathbf{\bar{y}}^\prime}\,\mathbf{\bar{y}}^\prime \right)}\,.
\end{eqnarray}
\renewcommand{\jot}{0em}
where
\begin{eqnarray} \label{eq:psiJ1}
\Delta_{\mathbf{\bar{y}}^\prime} &=& 
 \triangle(\graphr) \, \left[B(\graphr)\right]_i {\bar{r}'}_i 
+ \triangle(\graphgamma) \,  \left[B(\graphgamma)\right]_{(ij)} {\bar{\gamma }' }_{(ij)}
%
+\triangle(\graphrrra)
\, \left[B(\graphrrra)\right]_{i,j,k} {\bar{r}' }_i {\bar{r}' }_j {\bar{r}' }_k  
 \nonumber\\&&
 + \triangle(\graphgrrb) \, \left[B(\graphgrrb)\right]_{(ij),k,l} {\bar{\gamma }' }_{(ij)}{\bar{r}' }_k  {\bar{r}' }_l  
 +\left( \triangle(\graphggga) \,\left[B(\graphggga)\right]_{(ij),(kl),(mn)}+
\right.\nonumber\\&&\left.
  + \triangle(\graphgggb) \,\left[B(\graphgggb) \right]_{(ij),(kl),(mn)}
+  \triangle(\graphgggc) \, 
\left[B(\graphgggc)\right]_{(ij),(kl),(mn)}+
\right.\nonumber\\&&\left.
+ \triangle(\graphgggd) \,  \left[B(\graphgggd)\right]_{(ij),(kl),(mn)} + 
  \triangle(\graphggge) \, 
\left[B(\graphggge)\right]_{(ij),(kl),(mn)}
 \right.\nonumber\\&&\left.
 +  \triangle(\graphgggf) \, 
\left[B(\graphgggf) \right]_{(ij),(kl),(mn)}
 +  
\triangle(\graphgggg) \,  
\left[B(\graphgggg)\right]_{(ij),(kl),(mn)}  +
  \right.\nonumber\\&&\left.
  + \triangle(\graphgggh) \, \left[ B(\graphgggh) \right]_{(ij),(kl),(mn)}
\vphantom{\triangle(\graphrrrc)}   \right) {\bar{\gamma }' }_{(ij)} {\bar{\gamma }' }_{(kl)} {\bar{\gamma }' }_{(mn)}  
\\[1ex]
\left[ \mathbf{\bar{y}}^\prime\right]^T\,\mathbf{\bar{\Omega }}_{\mathbf{\bar{y}}^\prime}\,\mathbf{\bar{y}}^\prime
 &=&
 \left(   \triangle(\graphrra)  \, \left[B(\graphrra)\right]_{i,j} + \triangle(\graphrrb)  \, \left[B(\graphrrb)\right]_{i,j} \right) \, {\bar{r}' }_i {\bar{r}' }_j+
 \nonumber\\&&
+ \triangle(\graphgra) \, \left[ B(\graphgra)\right]_{(ij),k} {\bar{\gamma }' }_{(ij)}{\bar{r}' }_k       
  +  \left( \triangle(\graphgga) \, \left[B(\graphgga)\right]_{(ij),(kl)} 
  +
  \right.\nonumber\\&&\left.
  +\triangle(\graphggb) \,  \left[B(\graphggb)\right]_{(ij),(kl)} 
 +  \triangle(\graphggc) \, \left[B(\graphggc)\right]_{(ij),(kl)} 
\right)  {\bar{\gamma }' }_{(ij)} {\bar{\gamma }' }_{(kl)}\,.
\end{eqnarray}

%
Repeated indices $i$\,, $j$\,, \ldots imply summation from 1 to $N$\,, while repeated index pairs $(ij)$ etc imply the ordered sum $1 \leq i \leq j \leq N$\,. For example, $\left[B(\graphrrra)\right]_{i,j,k} {\bar{r}' }_i {\bar{r}' }_j {\bar{r}' }_k = \mbox{$\bar{r}'_1$}^3 + \mbox{$\bar{r}'_2$}^3 + \cdots \mbox{$\bar{r}'_N$}^3$\,.
In the above expressions for $\Delta_{\mathbf{\bar{y}}^\prime}$ and
$\left[ \mathbf{\bar{y}}^\prime\right]^T\,\mathbf{\bar{\Omega }}_{\mathbf{\bar{y}}^\prime}\,\mathbf{\bar{y}}^\prime$\,,
we are building up the invariant polynomials in ${\bar{r}'}_i$ and ${\bar{\gamma }' }_{(kl)}$ using the binary invariants as our building blocks.
 The scalar coefficients,
$\triangle(\mathcal{G})$ are (derived in the Appendix)
\begin{eqnarray}
\triangle(\graphrrra) & = &  \frac{ 1 }{3 \bar{r}^3_\infty}  \\
\triangle(\graphgrrb) & = &  \frac{\lambda-1}{N}  \\
\renewcommand{\jot}{0.5em}
\end{eqnarray}
\begin{eqnarray}
\triangle(\graphr) & = &  - \frac{1 }{\bar{r}_\infty}  \\
\triangle(\graphgamma) & = &  \mathcal{A}  \,  6\, (N+1) \gamma_\infty  \\
\triangle(\graphggga) & = &  \mathcal{A}  \, \mathcal{ \left( B + C\, D \right)  }   \\
\triangle(\graphgggb) & = &  \mathcal{A}  \,  \mathcal{  \left( B + C\, E + F \right)  }  \\
\triangle(\graphgggc) & = &  \mathcal{A}   \, \left( \mathcal{B} + \mathcal{C} \left( \frac{\mathcal{D}}{3} + \frac{ 2 \mathcal{E}}{3} \right) \right)  \\
\triangle(\graphgggd) & = &  \mathcal{A}  \, \mathcal{ \left( B + C\, E \right)  }   \\
\triangle(\graphggge) & = & \mathcal{A}  \, \left( \mathcal{B} + \frac{2 \mathcal{C E}}{3} -\mathcal{G} \right)  \\
\triangle(\graphgggf) & = & \mathcal{A}   \,   \left( \mathcal{B} + \frac{ \mathcal{C D}}{3} + 2\mathcal{G} \right)  \\
\triangle(\graphgggg) & = &  \mathcal{A}  \,  \left( \mathcal{B} + \frac{ \mathcal{C E}}{3} \right)  \\
\triangle(\graphgggh) & = & \mathcal{A}   \, \mathcal{B}  \\
\triangle(\graphrra) & = &  \lambda_\textrm{eff} + \frac{\lambda - 1}{2N}  (\lambda_\textrm{eff} -1) 
\end{eqnarray}
\begin{eqnarray}
\triangle(\graphrrb) & = &  \frac{\gamma_\infty}{2}  \\
\triangle(\graphgra) & = &  \bar{r}_\infty  \\
\triangle(\graphgga) & = &   \mathcal{H} \, ( \mathcal{I} + \mathcal{J} )   \\
\triangle(\graphggb) & = &   \mathcal{H} \, \left( \mathcal{I} - \frac{\gamma_\infty}{2} \right)   \\
\triangle(\graphggc) & = &    \mathcal{H}\,.
\end{eqnarray}
\renewcommand{\jot}{0.5em}
In the above equations, we have defined
\renewcommand{\jot}{0.5em}
\begin{eqnarray}
\lambda & = & \sqrt{1+N \lambda_p^2} \\
\lambda_p & = & \frac{\omega_p}{\omega_t} \\
\gamma_\infty & = & \frac{(\lambda -1)}{(N+(\lambda-1))}  \\
\bar{r}_\infty^2 & = & \frac{1}{2(1+(N-1)\gamma_\infty)} = \frac{N+(\lambda-1)}{2 \lambda  N}  \\
\lambda_\textrm{eff} & = &  \frac{N \lambda}{N+\lambda -1} 
\end{eqnarray}
\begin{eqnarray}
\mathcal{A} & = & \frac{1}{6 (1-\gamma_\infty) (1+(N-1) \gamma_\infty)} \\
\mathcal{B} & = &  - \frac{8 \gamma_\infty^3}{ (1-\gamma_\infty)^2 (1+(N-1) \gamma_\infty)^2 }  \\
\mathcal{C} & = &  - \frac{6 \gamma_\infty}{ (1-\gamma_\infty)^2 (1+(N-1) \gamma_\infty) }  \\
\mathcal{D} & = &    (1+(N-3) \gamma_\infty)  \\[-1ex]
\mathcal{E} & = &   - \gamma_\infty  \\
\mathcal{F} & = &   \frac{(1+(N-4) \gamma_\infty)}{(1-\gamma_\infty)^2}  \\
\mathcal{G} & = &  \frac{\gamma_\infty }{(1-\gamma_\infty)^2} \\
\mathcal{H} & = &   \frac{1}{ 2 (1-\gamma_\infty)^2 (1+(N-1) \gamma_\infty)}  \\
\mathcal{I} & = &  \frac{\gamma_\infty^2}{(1+(N-1) \gamma_\infty)}  \\
\mathcal{J} & = &   \frac{1+(N-3) \gamma_\infty}{2}\,.
\end{eqnarray}
\renewcommand{\jot}{0em}

This analytic solution through first order is then compared with the DPT
wavefunction derived from the general formalism of Paper~\onlinecite{wavefunction1} (see Eqs.~(33) and (117)) and implemented
in \textsc{Mathematica}\cite{mathematica} code. For the case of the general DPT formalism,
from Eqs.~(\ref{eq:groundWF}) and (\ref{eqphi0}) we find that
\begin{eqnarray}
 \mathbf{\bar{\Omega }}_{\mathbf{\bar{y}}^\prime}&=&\mathbf{V}^T\mathbf{\bar{\Omega }_{q'}}\mathbf{V} \,,
\\
 \mbox{} [\bar{\Omega }_{q'}]_{\nu_1,\nu_2}&=&\delta_{\nu_1,\nu_2}\bar{\omega}_{\nu_1} \,,
\end{eqnarray}
and $\mathbf{V}$ is the matrix transforming from the internal displacement coordinate vector $\mathbf{\bar{y}}^\prime$
to the normal mode coordinate vector $\mathbf{q'}$\,.
The polynomial $\Delta$ of Eq.~(\ref{eq:Phi1hat}) is similarly transformed from a normal coordinate basis to $\Delta_{\mathbf{\bar{y}}^\prime}$ of Eq.~(\ref{eq:wf1storder}) in the internal coordinate basis.

In Tables~\ref{tbl:st3diff}--\ref{tbl:wk3} we compare the binary invariant coefficients, $\triangle(\mathcal{G})$\,, from
the general theory of Papers~\onlinecite{wavefunction1} and Ref.~\onlinecite{normalmode} with the above results derived from the full analytical solution above for $N=10,000$ particles
and two different interparticle interaction strengths, $\lambda$\,. One value of $\lambda$ features strongly attractive harmonic interparticle interactions, while the other is for weakly-bound system with repulsive interparticle  interactions (negative $\lambda$)
for $\lambda$ just above the dissociation threshold at $\lambda = - 1/\sqrt{N}$\,.

 In both cases, to within round-off-error determined by the machine precision, exact agreement
is found, confirming the correctness of the general formalism of Paper~\onlinecite{wavefunction1}, and its implementation in
\textsc{Mathematica}\cite{mathematica} coding.

\section{Summary and Conclusions}
In this paper we performed the first test of a general 
formalism from Paper~\onlinecite{wavefunction1} for a fully-interacting
$N$-body wave function through first order in a perturbation
 expansion. This formalism was verified by
comparison to a  fully-interacting, exactly-solvable model problem.

The resources required for a solution to a general $N$-body problem
are understood to scale at least 
exponentially with $N$\,,
 making it very challenging to solve 
for large-$N$ systems.\cite{liu:2007,montina2008} 
The present perturbation series will scale exponentially in $N$ if summed to 
all orders.
However at  first order,   the number of terms scale as $N^6,$ a scaling, while greatly improved,  still remains 
challenging. Nonetheless, this $N^6$ scaling is tamed by 
expanding the perturbation series about a point where the $N$-body system has a
highly symmetric structure.  In the process, 
the $N$-scaling aspect also effectively separates away from the rest of the 
physics allowing the $N$ scaling to be treated as
a straight mathematical problem.

This highly symmetric structure for arbitrary $N$ is obtained as the number
of spatial dimensions approaches infinity, resulting in a configuration
 whose point group 
is isomorphic to the $S_N$ group.  All terms in the perturbation series for the
Hamiltonian are then invariant under the
$N!$ elements of the $S_N$ group, allowing an expansion in a basis that is also
invariant under these $N!$ operations.  This restriction results in a comparatively
small basis at each order which is independent of $N$. There are only seven
binary invariants at lowest order for any value of $N$, and twenty-three
 at next
order independent of $N$ (except when $N$ is quite small when the number is 
even lower). In this paper we demonstrated the completeness 
of this basis at all orders.  
Thus order-by-order the wave  function, along with other properties, may be
derived essentially analytically.

Since the perturbation parameter is the dimensionality of space and not 
the interaction,
this approach is equally applicable to weakly-interacting systems for 
which the mean-field approach is valid, and,
perhaps more interestingly, strongly-interacting systems for which the 
mean-field approach breaks down.
Paper~\onlinecite{wavefunction1} extends previous 
work\cite{GFpaper,energy,normalmode,density0} which derived energies, 
frequencies, normal-mode coordinates, wave functions,
and density profiles at lowest order for quantum systems of confined, 
interacting particles of any number, $N$\,. 


The general formalism set forth in Paper~\onlinecite{wavefunction1} for the wave function through first order, while
essentially analytic, has many moving parts, and so this paper sets out to verify the formalism by applying it to the
harmonically-confined system of $N$ particles interacting via harmonic potentials which may be attractive or repulsive.
This system is exactly soluble in $D$ dimensions, from which we directly derived the dimensional expansion
for the wave function through first order in terms of the binary invariants. This expansion, directly from the exact wave function,
has been compared with the wave function through first order from the general formalism of Paper~\onlinecite{wavefunction1}. Since at each order
there  are only a finite number of binary invariants to consider, there are only a
finite number of coefficients to the binary invariant terms that have to be compared. Exact agreement is found between the
coefficients obtained directly from the analytic solution and those derived using the general formalism of
Paper~\onlinecite{wavefunction1}, confirming this general formalism.

The formalism of Paper~\onlinecite{wavefunction1} is not limited to the harmonic interactions
discussed in this paper, and indeed is quite general. In previous papers we
have examined other potentials, in particular the hard-sphere potential in relation to Bose-Einstein condensates, at lowest order.
While the lowest-order formalism adequately captures the behavior of the system in a range of scattering length, $a$\,,
and particle number $N$ at which the mean field fails, for large enough $a$ and/or $N$ the lowest-order wave function
no longer has the flexibility to adequately represent the actual system.
It is thus desirable to apply the general formalism of Paper~\onlinecite{wavefunction1} for the wave function,
through first-order and verified in this paper, to other systems
such as the Bose-Einstein condensate for interaction strengths and particle number at which the mean field breaks down.

In principle any observable quantity can be obtained from the wave function and as an illustration in Ref.~\onlinecite{density0} we derived the
density profile at lowest order from the lowest-order wave function. With the next-order wave function available from the formalism
of Paper~\onlinecite{wavefunction1} we can now derive any observable quantity, such as the density profile, to next order in the perturbation theory.

It is also important to note that while Paper~\onlinecite{wavefunction1} derives the $N$-particle wave function to next order in perturbation theory, the same
basic approach can, in principle, be used to derive yet higher-order terms in the perturbation series.


\appendix
\section{Confined, Harmonically Interacting, Analytically Solvable Model System}
\label{sec:HarmIntWF}
In this appendix we derive the exact ground-state wave function for a harmonically-confined, harmonically-interacting system of
$N$ particles in $D$ dimensions, and from it derive the wave function through first order in $\delta^{1/2}$ exactly, where
$\delta = 1/D$\,.

The Hamiltonian of the harmonically interacting model system of identical particles is
\begin{equation}
H = \frac{1}{2} \, \left( \sum_i^N  \left[ - \frac{\partial^2}{\partial \bm{r}_i^2} + \omega_t^2 \bm{r}_i^2 \right]  + \sum_{i<j}^{N} \omega_p^2 \bm{r}_{ij}^2 \right) \,.
\end{equation}
Making the orthogonal transformation to center of mass and Jacobi coordinates
\begin{eqnarray}
\bm{R} = \frac{1}{\sqrt{N}} \sum_{k=1}^N \bm{r}_k & \hspace{1ex} \mbox{and} \hspace{1ex} & \bm{\rho}_i = \frac{1}{\sqrt{i(i+1)}} \left( \sum_{j=1}^i \bm{r}_j- i \bm{r}_{i+1} \right) \,,
\end{eqnarray}
where $1 \leq i \leq N-1$\,, the Hamiltonian becomes
\begin{equation}
H = \frac{1}{2} \, \left( - \frac{\partial^2}{\partial \bm{R}^2} + \omega_t^2 \bm{R}^2 \right)
+ \frac{1}{2} \, \sum_{i=1}^{N-1} \left( - \frac{\partial^2}{\partial \bm{\rho }_i^2} + \omega_{\rm int}^2 \bm{\rho}^2 \right) \,,
\end{equation}
the sum of $N$\,, $D$-dimensional harmonic-oscillator Hamiltonians, where
\begin{equation}
\omega_{\rm int} = \sqrt{\omega_t^2 + N\, \omega_p^2}\,.
\end{equation}

Notice two things about the Hamiltonian: it is separable and each component has the form of a $D$-dimensional harmonic oscillator. Therefore the ground-state solution to the wave function in the Schr\"odinger equation
\begin{equation}
 H \, \Psi = E \, \Psi
\end{equation}
is the product of harmonic-oscillator wavefunctions
\begin{equation}
\label{eq:WFNJ}
\Psi(\bm{R}, \, \{\bm{\rho}_i\}; \, D) = \psi(R;\,  \omega_t, \, D) \, \prod_{i=1}^{N-1} \psi(\rho_i; \, \omega_{\rm int }, \, D) \,,
\end{equation}
where $\psi(\rho_i; \, \omega_{\rm int }, \, D)$ is the $D$-dimensional, harmonic-oscillator, ground-state wave function
\begin{equation}
\label{eq:DHarmWF}
\psi(r; \, \omega, \, D) = \sqrt{\frac{2 \omega^{\frac{D}{2}}}{\Gamma(\frac{D}{2})}} \,\, \exp{\left( - \frac{\omega}{2} r^2 \right) }
\end{equation}
satisfying the normalization condition
\begin{equation}
\int_0^\infty [\psi(r; \, \omega, \, D)]^2 \,r^{D-1}\,dr = 1 \,.
\end{equation}

The Jacobian-weighted, $L=0$ wave function $\Psi_J$ is obtained by folding into the wavefunction, the square root of that portion of the Jacobian which depends on the internal coordinates, i.e.\ the square root of
\begin{equation}
\Gamma^{(D-N-1)/2} \prod_{j=1}^N r_j^{(D-1)}\,,
\end{equation}
where $\Gamma$ is the Grammian determinant, so that
\begin{equation} \label{eq:PsiJ}
\Psi_J = \mathcal{N} \, \Gamma^{(D-N-1)/4} \prod_{j=1}^N r_j^{(D-1)/2} \, \psi(R;\,  \omega_t, \, D) \, \prod_{i=1}^{N-1} \psi(\rho_i; \, \omega_{\rm int }, \, D) \,,
\end{equation}
where $\mathcal{N}$ is a normalization constant ensuring that
\begin{equation}
\int [\Psi_J(\{r_i\}, \, \{\gamma_{jl} \}; D)]^2 \prod_i dr_i \prod_{j < k} d\gamma_{jk} = 1 \,.
\end{equation}
\subsection{A perturbation series in $1/\sqrt{D}$ for the exact wavefunction}
\subsubsection{Dimensional scaling}
Now consider transforming to dimensionally-scaled oscillator coordinates
\renewcommand{\arraystretch}{1.5}
\begin{equation}
\begin{array}{rcl@{\hspace{2em}}c@{\hspace{2em}}rcl@{\hspace{2em}}c@{\hspace{2em}}rcl}
r & = & D^2 \, \bar{a}_t \bar{r} & & \rho & = & D^2 \, \bar{a}_t \bar{\rho} & & R & = & D^2 \, \bar{a}_t \bar{R} \\
 & = & {\displaystyle \frac{D^2 \, \bar{r}}{\sqrt{\bar{\omega}_t}}} & &  & = & {\displaystyle \frac{D^2 \, \bar{\rho }}{\sqrt{\bar{\omega}_t}}} & &  & = & {\displaystyle \frac{D^2 \, \bar{R}}{\sqrt{\bar{\omega}_t}}} \\
 & = & {\displaystyle \frac{D^{1/2} \, \bar{r}}{\sqrt{\omega_t}}} & &  & = & {\displaystyle \frac{D^{1/2} \, \bar{\rho }}{\sqrt{\omega_t}}}
&  &  & = & {\displaystyle \frac{D^{1/2} \, \bar{R}}{\sqrt{\omega_t}}} \,, \end{array}
\end{equation}
\renewcommand{\arraystretch}{1}
where the $\omega_t = \frac{\bar{\omega}_t}{D^3}$\,, and $\bar{a}_t$ is the dimensionally-scaled oscillator length of the trap. Both $\bar{r}$ and $\bar{\rho}$ are dimensionless. From Eq.~(\ref{eq:PsiJ}) we obtain
\begin{eqnarray}
\label{eq:all_order_Jacobian-weighted_Psi}
\Psi_J &=& \mathcal{N} \, \Gamma^{(D-N-1)/4} \prod_{j=1}^N \bar{r}_j^{(D-1)/2} \,
\sqrt{ \frac{2}{\Gamma(\frac{D}{2} ) } } \, D^{\frac{D}{4}} \, \exp{\left(- \frac{D}{2} \, \bar{R}^2 \right)} \nonumber \\ && \times
\left( \frac{2}{\Gamma(\frac{D}{2} ) } \right)^{\frac{N-1}{2} } \, [(\lambda D)^{\frac{D}{2}}]^{\frac{N-1}{2} }
 \, \prod_{i=1}^{N-1} \exp{\left(- \frac{\lambda D}{2} \, \bar{\rho }^2_i \right)} \,, \label{eq:psiJ}
\end{eqnarray}
where
\begin{equation}
\lambda =\frac{\omega_{\rm int}}{\omega_t} \,.
\end{equation}
\subsubsection{The large-dimension limit}
To test the general formalism of Paper~\onlinecite{wavefunction1} we need to expand Eq.~(\ref{eq:all_order_Jacobian-weighted_Psi})
about the large-dimension limit through first order in $\delta^{1/2}$\,.
In the large-dimension limit the system localizes about a structure where all the radii are equal to $\bar{r}_\infty$ and angle cosines are equal to $\gamma_\infty$. To derive $\bar{r}_\infty$ and $\gamma_\infty$ one applies the condition
\begin{equation} \label{eq:wfmax}
\left. \frac{\partial \Psi_J }{\partial \bar{r}_i } \right|_{D=\infty} = \left. \frac{\partial \Psi_J }{\partial \gamma_{jk} } \right|_{D=\infty} = 0 \,.
\end{equation}
In this endeavor the following results are useful:
\renewcommand{\jot}{0.5em}
\begin{eqnarray}
\left. \Gamma \right|_{D=\infty} & = & (1+(N-1)\gamma_\infty) (1-\gamma_\infty)^{N-1} \label{eq:GammaInf0} \\
\left. \frac{\partial \Gamma}{\partial \gamma_{jk}} \right|_{D=\infty} & = & -2 \gamma_\infty (1-\gamma_\infty)^{N-2} \label{eq:GammaInf1} \\
\left. \bar{R}^2 \right|_{D=\infty} & = & \bar{r}_\infty^2 (1+(N-1) \gamma_\infty) \\
\left. \frac{\partial \bar{R}^2}{\partial \bar{r}_i} \right|_{D=\infty} & = & 2 \bar{r}_\infty  \frac{(1+(N-1) \gamma_\infty)}{N} \\
\left. \frac{\partial \bar{R}^2}{\partial \gamma_{jk}} \right|_{D=\infty} & = & 2 \, \frac{ \bar{r}_\infty^2}{N} \\
\sum_{i=1}^{N-1} \bar{\rho}_i^2 & = & \sum_{j=1}^N \bar{\rho}_j^2 - \bar{R}^2 \\
\left. \sum_{i=1}^{N-1} \bar{\rho}_i^2 \, \right|_{D=\infty} & = &  (N-1) \bar{r}_\infty^2 (1-\gamma_\infty) = (N-1) \bar{\rho}_\infty \\
\left. \frac{ \partial \sum_{i=1}^{N-1} \bar{\rho}_i^2 }{\partial r_k} \right|_{D=\infty} & = & \frac{2 (N-1)}{N} \bar{r}_\infty (1-\gamma_\infty) \\
\left. \frac{ \partial \sum_{i=1}^{N-1} \bar{\rho}_i^2 }{\partial \gamma_{jk} } \right|_{D=\infty} & = & - \frac{2}{N} \bar{r}_\infty^2\,.
\end{eqnarray}
\renewcommand{\jot}{0em}
>From Eq.~(\ref{eq:wfmax}) we obtain the parameters $\bar{r}_\infty$ and $\gamma_\infty$ 
\begin{eqnarray}
\gamma_\infty & = & \frac{(\lambda -1)}{(N+(\lambda-1))}  \label{eq:ginf} \\
\bar{r}_\infty^2 & = & \frac{1}{2(1+(N-1)\gamma_\infty)} = \frac{N+(\lambda-1)}{2 \lambda  N}  \label{eq:rinf}  \,.
\end{eqnarray}
Equations~(\ref{eq:ginf}) and (\ref{eq:rinf}) define the $D \rightarrow \infty$ structure about which the system oscillates at finite dimension.

\subsubsection{A series expansion about the large-$D$ limit}

To derive the wave function through order $\delta^{1/2}$ we perform a series expansion of each of the $D$-dependent terms in Eq.~(\ref{eq:all_order_Jacobian-weighted_Psi}). 
%
\begin{equation} \label{eq:sqrtoneOGamma}
\sqrt{\frac{1}{\Gamma\left(\frac{D}{2}\right)} } = \frac{2^{\frac{D-2}{4}}\exp{(\frac{D}{4}}) }{\sqrt[4]{\pi} D^{\frac{D-1}{4}} }
+ O(\delta) \,,
\end{equation}
\begin{equation} \label{eq:prodr}
\renewcommand{\arraystretch}{2}
\begin{array}{rcl}
{\displaystyle \prod_{i=1}^N \bar{r}_i^{\frac{D-1}{2}} = } & & {\displaystyle \bar{r}_\infty^{\frac{N(D-1)}{2}} \exp{ \left(  \sum_{i=1}^N
\frac{D^\frac{1}{2} \bar{r}'_i }{2 \bar{r}_\infty}    \right) } \, \exp{ \left(  - \frac{1}{4 \bar{r}^2_\infty} \sum_{i=1}^N
\bar{r}^{\prime \, 2}_i \right)  }  }  \\
& \times & {\displaystyle \left( 1 + \frac{ \delta^{ \frac{1}{2}} } {2} \sum_{i=1}^N \left(  
\frac{\bar{r}^{\prime \, 3}_i }{3 \bar{r}^3_\infty} - \frac{\bar{r}'_i}{\bar{r}_\infty}  \right)  + O(\delta)     \right)   \,. }
\end{array}
\renewcommand{\arraystretch}{1}
\end{equation}
We also have
\renewcommand{\jot}{0.5em}
\begin{eqnarray}
\bar{R}^2  & = & \bar{R}^2_\infty + \delta^{1/2} \, \bar{R}_2^{\prime}(\delta^{1/2}) \\
\sum_{i=1}^{N-1} \bar{\rho }^2_i  & = & (N-1) \, \bar{\rho }^2_\infty + \delta^{1/2} \, _{{}_{\sum}}\bar{\rho }_2^{\prime}(\delta^{1/2})\,,
\end{eqnarray}
\renewcommand{\jot}{0em}
where
\renewcommand{\jot}{0.5em}
\begin{eqnarray}
\left. \bar{R}^2 \right|_{D \rightarrow \infty}  & \equiv & \bar{R}^2_\infty = \bar{r}_\infty^2 (1+(N-1) \gamma_\infty) \\
\left. \sum_{i=1}^{N-1} \bar{\rho }^2_i \right|_{D \rightarrow \infty} & \equiv & (N-1) \, \bar{\rho }^2_\infty  =
 (N-1) \, \bar{r}_\infty^2 (1- \gamma_\infty)
\end{eqnarray}
\renewcommand{\jot}{0em}
and
\renewcommand{\jot}{0.5em}
\begin{eqnarray}
\bar{R}_2^{\prime}(\delta^{1/2})  & = &   \begin{array}[t]{rl} & {\displaystyle \frac{2 \bar{r}_\infty}{N}
\left( (1+(N-1) \gamma_\infty) \sum_{i=1}^N \bar{r}'_i + \sum_{i<j=1}^N \bar{r}_\infty \bar{\gamma}'_{ij}\right) }\\
+ & {\displaystyle\frac{\delta^{1/2}}{N} \left(  \sum_{i=1}^N (\bar{r}'_i )^2
+ 2 \gamma_\infty \!\! \sum_{i<j=1}^N \bar{r}'_i \bar{r}'_j   +2 \bar{r}_\infty \!\! \sum_{i<j=1}^N (\bar{r}'_i + \bar{r}'_j) \bar{\gamma}'_{ij} \right)   } \\
+ & {\displaystyle \delta \frac{2}{N} \sum_{i<j=1}^N \bar{r}'_i \bar{r}'_j \bar{\gamma}'_{ij} }
\end{array} \\
_{{}_{\sum}}\bar{\rho }_2^{\prime}(\delta^{1/2})  & = &  \begin{array}[t]{rl} & {\displaystyle \frac{2 \bar{r}_\infty}{N}
\left( (N-1) (1-\gamma_\infty) \sum_{i=1}^N \bar{r}'_i - \sum_{i<j=1}^N \bar{r}_\infty \bar{\gamma}'_{ij}\right) }\\
+ & {\displaystyle\frac{\delta^{1/2}}{N} \left(  (N-1) \sum_{i=1}^N (\bar{r}'_i )^2
- 2 \gamma_\infty \!\! \sum_{i<j=1}^N \bar{r}'_i \bar{r}'_j -  2 \bar{r}_\infty \!\! \sum_{i<j=1}^N (\bar{r}'_i + \bar{r}'_j) \bar{\gamma}'_{ij} \right)   } \\
- & {\displaystyle \delta \frac{2}{N} \sum_{i<j=1}^N \bar{r}'_i \bar{r}'_j \bar{\gamma}'_{ij} } \,,
\end{array}
\end{eqnarray}
\renewcommand{\jot}{0em}
so that
\renewcommand{\jot}{0.5em}
\begin{eqnarray}
\lefteqn{\exp{\left(- \frac{D}{2} \, \bar{R}^2 \right)} \, \prod_{i=1}^{N-1} \exp{\left(- \frac{\lambda D}{2} \, \bar{\rho }^2_i \right)}} \nonumber \\
& = &  \renewcommand{\arraystretch}{2}\begin{array}[t]{@{}rl} & {\displaystyle \exp{ \left( - \frac{D\, N}{4}   \right) } \, \exp{ \left(  - \sum_{i=1}^N
\frac{D^\frac{1}{2} \bar{r}'_i }{2 \bar{r}_\infty}    \right) } \, 
\exp{ \left(  - \frac{D^{\frac{1}{2}} \bar{r}^2_\infty}{N} \, (1-\lambda) \sum_{i<j=1}^N \bar{\gamma}'_{ij}  \right) } }  \\
  \times & {\displaystyle \exp{ \left(  - \frac{1}{2} \left(  \vphantom{\sum_{i<j=1}^N} \right. \right. } \renewcommand{\arraystretch}{2} \begin{array}[t]{@{}l}  
    {\displaystyle    \left( \lambda - \frac{\lambda-1}{N} \right) \sum_{i=1}^N \bar{r}_i^{\prime \, 2}   
 -\frac{2(\lambda-1)}{N} \, \bar{\gamma}_\infty \!\! \sum_{i<j=1}^N \bar{r}'_i \bar{r}'_j  }   \\
 {\displaystyle \left. \left. -\frac{2(\lambda-1)}{N} \, \bar{r}_\infty \!\! \sum_{i<j=1}^N (\bar{r}'_i  + \bar{r}'_j) \bar{\gamma}'_{ij} \right)    \right) }   \,
\end{array} } \\
 \times & {\displaystyle \left(  1 + \delta^{\frac{1}{2}} \left( \frac{\lambda -1}{N} \right) \sum_{i<j=1}^N \bar{r}'_i \bar{r}'_j \bar{\gamma }'_{ij}  + O(\delta) \right) }
\end{array} \label{eq:expRexprho}\,.
\end{eqnarray}
\renewcommand{\arraystretch}{1}
\renewcommand{\jot}{0em}
The final bit of the puzzle in the dimensional expansion of Eq.~(\ref{eq:psiJ}) is the dimensional expansion
of $\Gamma^{(D-N-1)/4}$\,. For this we need Eqs.~(\ref{eq:GammaInf0}), (\ref{eq:GammaInf1}), and
\renewcommand{\jot}{0.5em}
\begin{eqnarray}
\left. \frac{\partial^2 \Gamma}{\partial \gamma_{ij} \, \partial \gamma_{kl}} \right|_{D=\infty} & = & 0 \\
\left. \frac{\partial^2 \Gamma}{\partial \gamma_{ij} \, \partial \gamma_{jk}} \right|_{D=\infty} & = & 2 \gamma_\infty (1-\gamma_\infty)^{N-3} \\
\left. \frac{\partial^2 \Gamma}{\partial \gamma_{ij}^2} \right|_{D=\infty} & = & -2 (1+(N-3)\gamma_\infty) (1-\gamma_\infty)^{N-3} \\
\left. \frac{\partial^3 \Gamma}{\partial \gamma_{ij} \, \partial \gamma_{kl} \, \partial \gamma_{mn} } \right|_{D=\infty} & = & 0 \\
\left. \frac{\partial^3 \Gamma}{\partial \gamma_{ij} \, \partial \gamma_{jk} \, \partial \gamma_{lm} } \right|_{D=\infty} & = & 0 \\
\left. \frac{\partial^3 \Gamma}{\partial \gamma_{ij} \, \partial \gamma_{jk} \, \partial \gamma_{kl} } \right|_{D=\infty} & = & - 2 \gamma_\infty (1-\gamma_\infty)^{N-4} \\
\left. \frac{\partial^3 \Gamma}{\partial \gamma_{ij} \, \partial \gamma_{jk} \, \partial \gamma_{jl} } \right|_{D=\infty} & = & 0 \\
\left. \frac{\partial^3 \Gamma}{\partial \gamma_{ij} \, \partial \gamma_{jk} \, \partial \gamma_{ik} } \right|_{D=\infty} & = & 2 (1+(N-4)\gamma_\infty) (1-\gamma_\infty)^{N-4} \\
\left. \frac{\partial^3 \Gamma}{\partial \gamma_{ij}^2 \, \partial \gamma_{kl} } \right|_{D=\infty} & = & 4 \gamma_\infty (1-\gamma_\infty)^{N-4} \\
\left. \frac{\partial^3 \Gamma}{\partial \gamma_{ij}^2 \, \partial \gamma_{jk} } \right|_{D=\infty} & = & 0 \\
\left. \frac{\partial^3 \Gamma}{\partial \gamma_{ij}^3 } \right|_{D=\infty} & = & 0 \,,
\end{eqnarray}
\renewcommand{\jot}{0em}
from which we obtain
\renewcommand{\jot}{0.5em}
\begin{eqnarray} \lefteqn{
\Gamma^{(D-N-1)/4} = \left( (1-\gamma_\infty)^{N-1} \, (1+(N-1) \gamma_\infty) \right)^{\frac{D-N-1}{4} } }  \nonumber \\
&& \times \left( 1 + \frac{\delta^{\frac{1}{2}}}{12 (1-\gamma_\infty)(1+(N-1)\gamma_\infty)}
\left( - \frac{8 \gamma_\infty^3}{ (1-\gamma_\infty)^2(1+(N-1)\gamma_\infty)^2} [ B(\graphgamma) \bm{\bar{\gamma }' } ]^3 \right. \right. \nonumber \\
&& - \frac{6 \gamma_\infty}{ (1-\gamma_\infty)^2(1+(N-1)\gamma_\infty)} [ B(\graphgamma) \bm{\bar{\gamma }' } ]
\{ (1+(N-3)\gamma_\infty) B(\graphgga) - \gamma_\infty B(\graphggb) \}  \bm{\bar{\gamma }' } \bm{\bar{\gamma }' } \hphantom{xxxxxx}
\nonumber \\
&& + \frac{1}{(1-\gamma_\infty)^2} \{(1+(N-4) \gamma_\infty) B(\graphgggb) - \gamma_\infty B(\graphggge) + 2 \gamma_\infty B(\graphgggf) \}  \bm{\bar{\gamma }' } \bm{\bar{\gamma }' } \bm{\bar{\gamma }' }
\nonumber \\
&& \left. \left. + \, 6(N+1) \gamma_\infty [ B(\graphgamma) \bm{\bar{\gamma }' } ]  \vphantom{\frac{8 \gamma_\infty^3}{ (1-\gamma_\infty)^2(1+(N-1)\gamma_\infty)^2}}  \right) + O(\delta)  \vphantom{\frac{\delta^{\frac{1}{2}}}{12 (1-\gamma_\infty)(1+(N-1)\gamma_\infty)} }   \right) 
\,\, \exp{\left( - D^{\frac{1}{2}}  \, \frac{(\lambda-1) \bar{r}_\infty^2 }{N} \sum_{i <  j = 1}^N \bar{\gamma}'_{ij}
\right) }  \nonumber \\
&& \times \exp{\left(-\frac{1}{2 (1-\gamma_\infty)^2 (1+(N-1)\gamma_\infty)} \left( - \frac{ \gamma_\infty^2}{ (1+(N-1)\gamma_\infty)} [ B(\graphgamma) \bm{\bar{\gamma }' } ]^2 \right. \right.   }  \nonumber \\
&& \left. \left. + \left[ \frac{ (1+(N-3)\gamma_\infty ) }{  2 }  B(\graphgga) - \frac{\gamma_\infty}{2} B(\graphggb)
\right] \bm{\bar{\gamma }' } \bm{\bar{\gamma }' }
\right)
\right) \,, \label{eq:gamma_exp}
\end{eqnarray}
\renewcommand{\jot}{0em}
where the $[B ({\mathcal G})]_{\nu_1, \nu_2, \ldots}$ are the binary invariants introduced in Paper~\onlinecite{wavefunction1} (briefly
reviewed in Appendix~\ref{app:Bin}) and $\mathcal{G}$ is the graph labeling the binary invariant.
The expression  $B(\mathcal{G}) \bm{\bar{X}'_1 } \bm{\bar{X_2 }' } \bm{\bar{X_3}' }$ is shorthand for
$[B(\mathcal{G})]_{\nu_1, \nu_2, \nu_3} \, [\bm{\bar{X}'_1 }]_{\nu_1} [\bm{\bar{X_2 }' }]_{\nu_2} [\bm{\bar{X_3}' }]_{\nu_2}$ where
repeated indices $\nu_i$ are summed over,
$\bm{\bar{X }' }$ is the $\bm{\bar{r }' }$ or $\bm{\bar{\gamma }' }$ vector from Eq.~(\ref{eq:ytransposeP}), likewise for
$B(\mathcal{G}) \bm{\bar{X}'_1 } \bm{\bar{X_2 }' }$ and $B(\mathcal{G}) \bm{\bar{X}'_1 }$\,.
Using Eqs.~(\ref{eq:sqrtoneOGamma}), (\ref{eq:prodr}), (\ref{eq:expRexprho}), and (\ref{eq:gamma_exp}),
along with
\begin{eqnarray}
B(\graphgamma) \otimes B(\graphgamma) & = & B(\graphgga) + B(\graphggb) + B(\graphggc)  \\
B(\graphgamma) \otimes B(\graphgamma) \otimes B(\graphgamma) & = & B(\graphggga) + B(\graphgggb) + B(\graphgggc) + B(\graphgggd)
+ B(\graphggge) ) \nonumber \\ && + B(\graphgggf) + B(\graphgggg) + B(\graphgggh) \\[1.5ex]
B(\graphgamma) \otimes B(\graphgga) & = & B(\graphggga) + \frac{B(\graphgggc)}{3} + \frac{B(\graphgggf)}{3} \\
B(\graphgamma) \otimes \frac{B(\graphggb)}{2} & = & \frac{B(\graphgggc)}{3} + \frac{B(\graphgggb)}{2}
+ \frac{B(\graphgggd)}{2} + \frac{B(\graphggge)}{3}  + \frac{B(\graphgggg ) \vphantom{\begin{array}[t]{c}
x \end{array} } }{6}
\end{eqnarray}
in Eq.~(\ref{eq:psiJ}), with
\begin{equation}
\mathcal{N} = \frac{1}{ \bar{r}_\infty^{\frac{N(D-1)}{2}} \, \left( (1-\gamma_\infty)^{N-1} \, (1+(N-1) \gamma_\infty) \right)^{\frac{D-N-1}{4} } } + O(\delta)  \,,
\end{equation}
we obtain the Jacobian-weighted $N$-body wavefunction in Eq.~(\ref{eq:psiJ1}) for a system of identical particles under harmonic confinement with harmonic interactions.


\begin{table}[p]\centering
\begin{tabular}{|c|c|} \hline
$\mathcal{G}$&$\Delta\left[\triangle(\mathcal{G})\right]$\\
\hline
$\graphrrra$ &$7.0\times 10^{-16}$\\ 
$\graphgrrb$ &$2.3\times 10^{-11}$\\
$\graphggga$ &$-4.1\times 10^{-16}$\\ 
$\graphgggb$ &$-1.3\times 10^{-16}$\\ 
$\graphgggc$ &$-6.1\times 10^{-16} $\\ $\graphgggd$&$1.2\times 10^{-16}$
\\ \hline 
\end{tabular}
\begin{tabular}{|c|c|} \hline
$\mathcal{G}$&$\Delta\left[\triangle(\mathcal{G})\right]$\\
\hline
 $\graphggge$ &$-1.5\times 10^{-16} $ \\
 $\graphgggf$ &$-6.0\times 10^{-13} $\\
$\graphgggg$ &$ 3.7\times 10^{-16}$ \\
$\graphgggh$ & $ -2.1\times 10^{-13}$\\
$\graphr$ & $ -8.4\times 10^{-7} $\\
$\graphgamma$ & $8.4\times 10^{-10}$
\\ \hline 
\end{tabular}
\begin{tabular}{|c|c|} \hline
$\mathcal{G}$&$\Delta\left[\triangle(\mathcal{G})\right]$\\
\hline
$\graphrra$ & $-3.6\times 10^{-16} $\\
$\graphrrb$ &$-5.1\times 10^{-11} $\\
$\graphgra$ & $9.6\times 10^{-15}$\\
$\graphgga$ & $-3.7\times 10^{-16} $\\
$\graphggb$ & $-2.2\times 10^{-16} $\\
$\graphggc$&$ 2.5\times 10^{-14}$
\\ \hline
\end{tabular}
\caption{Fractional difference,  $\Delta\triangle(\mathcal{G}) =( \triangle_{analytic}(\mathcal{G}) -\triangle_{DPT}(\mathcal{G}) ) /\triangle_{analytic}(\mathcal{G}) $\,, between the analytic and DPT rank-three, rank-two, and rank-one binary invariant coefficients when
$N=10,000$ and $\lambda = 10$.}
\label{tbl:st3diff}
\end{table}

\begin{table}[p]
\centering
\begin{tabular}{|c|c|} \hline
$\mathcal{G}$&$\triangle(\mathcal{G})$\\
\hline
$\graphrrrb$ & $2.3\times 10^{-11} $\\
$\graphrrrc$ & $ 1.1\times 10^{-11}$\\
$\graphgrra$ & $ 2.4\times 10^{-15} $\\
$\graphgrrc$ & $ 2.2\times 10^{-15}$\\
$\graphgrrd$ & $ 2.0\times 10^{-15}$\\
$\graphgrre$ &$ 8.2\times 10^{-16}$ \\
 $\graphgrb$ &$
-1.2\times 10^{-17}$
\\
\hline
\end{tabular}
\begin{tabular}{|c|c|} \hline
$\mathcal{G}$&$\triangle(\mathcal{G})$\\
\hline
$\graphggra$ & $-2.9\times 10^{-15} $\\
$\graphggrb$ & $ 7.8\times 10^{-19} $\\
$\graphggrc$ & $4.1\times 10^{-19} $\\
$\graphggrd$ & $9.8\times 10^{-17} $\\
$\graphggre$  & $ 1.0\times 10^{-19}$\\
$\graphggrf$  & $6.2\times 10^{-20} $\\
$\graphggrg$ & $2.1\times 10^{-20}$
\\ \hline
\end{tabular}
\caption{Rank-three and rank-two binary invariant coefficient,  $\triangle_{DPT}(\mathcal{G}) $\,, from the general \emph{Mathematica} code when
$N=10,000$ and $\lambda = 10$. All of these coefficients are exactly zero in the exactly soluble analytic  solution.}
\label{tbl:st3}
\end{table}


\begin{table}[p]\centering
\begin{tabular}{|c|c|} 
\hline
$\mathcal{G}$&$\Delta\left[\triangle(\mathcal{G})\right]$\\
\hline
$\graphrrra$ & $ 1.6\times 10^{-11}$\\
$\graphgrrb$ &$-3.8\times 10^{-8} $\\
$\graphggga$ & $2.5\times 10^{-8} $\\
$\graphgggb$ & $-3.8\times 10^{-8} $\\
$\graphgggc$ & $2.5\times 10^{-8} $\\
$\graphgggd$&$ 2.2\times 10^{-5}$
\\ \hline 
\end{tabular}
\begin{tabular}{|c|c|} \hline
$\mathcal{G}$&$\Delta\left[\triangle(\mathcal{G})\right]$\\
\hline
 $\graphggge$ & $-2.0\times 10^{-8}$\\
 $\graphgggf$ & $ 1.7\times 10^{-13}$\\
$\graphgggg$ & $ 2.1\times 10^{-11} $\\
$\graphgggh$ & $ -3.0\times 10^{-14}$\\
 $\graphr$ & $1.7\times 10^{-3} $\\
 $\graphgamma$&$-1.7\times 10^{-10}$
\\ \hline
\end{tabular}
\begin{tabular}{|c|c|} \hline
$\mathcal{G}$&$\Delta\left[\triangle(\mathcal{G})\right]$\\
\hline
$\graphrra$ &$-2.6\times 10^{-11} $\\
$\graphrrb$ &$ -1.1\times 10^{-9}$\\
$\graphgra$ & $6.8\times 10^{-13}$\\
$\graphgga$ & $-1.6\times 10^{-7}$\\
$\graphggb$ & $ 1.6\times 10^{-10}$\\
$\graphggc$&$ -3.8 \times 10^{-13}$
\\ \hline
\end{tabular}
\caption{Fractional difference,  $\Delta\triangle(\mathcal{G}) =( \triangle_{analytic}(\mathcal{G}) -\triangle_{DPT}(\mathcal{G}) ) /\triangle_{analytic}(\mathcal{G}) $\,, between the analytic and DPT rank-three, rank-two and rank-one binary invariant coefficients when
$N=10,000$ and $\lambda^2 = -1/10,000 + 10^{-10}$.}
\label{tbl:wk3diff}
\end{table}
\begin{table}[p]\centering
\renewcommand\extrarowheight{1ex}
\begin{tabular}{|c|c|} \hline
$\mathcal{G}$&$\triangle(\mathcal{G})$\\
\hline
$\graphrrrb$ &$-4.8\times 10^{-20} $\\ 
$\graphrrrc$ &$ 7.9\times 10^{-24}$\\ 
$\graphgrra$ &$1.9\times10^{-12}$\\ 
$\graphgrrc$ &$-1.9\times10^{-16}$\\
 $\graphgrrd$  &$-2.6\times10^{-19}$\\ 
 $\graphgrre$ &$3.8\times10^{-20}$\\
$\graphgrb$&$8.1\times 10^{-20}$
\\
\hline 
\end{tabular}
\begin{tabular}{|c|c|} \hline
$\mathcal{G}$&$\triangle(\mathcal{G})$\\
\hline
$\graphggra$ &$2.8\times 10^{-10}$\\ 
$\graphggrb$ &$ -1.7\times 10^{-14}$\\
 $\graphggrc$ &$ -2.8\times 10^{-14}$ \\
  $\graphggrd$ &$5.3\times 10^{-14} $\\ 
  $\graphggre$  & $-1.2\times 10^{-17}$\\ 
  $\graphggrf$  &$ -5.0\times 10^{-18} $\\ $\graphggrg$&$3.3\times 10^{-20}$
\\ \hline
\end{tabular}
\caption{Rank-three and rank-two binary invariant coefficient,  $\triangle_{DPT}(\mathcal{G}) $\,, from the general \emph{Mathematica} code when
$N=10,000$ and $\lambda^2 = -1/10,000 + 10^{-10}$. All of these coefficients are exactly zero in the exactly soluble analytic  solution.}
\label{tbl:wk3}
\end{table}


\begin{thebibliography}{99}
\bibitem{wavefunction1} W.~Laing, M.~Dunn, and D.~Watson,
\newblock Arxiv preprint math-ph/0808.2949v1; J.\ Math.\ Phys., submitted .
\bibitem{liu:2007}
Y.~Liu, M.~Christandl, and F.~Verstraete,
\newblock Phys.\ Rev.\ Lett.\ {\bf 98}, 110503 (2007).
\bibitem{montina2008}
A.~Montina,
\newblock Phys.\ Rev.\ A {\bf 77}, 22104 (2008).
\bibitem{MKR}
D.~Masiello, S.~McKagan, and W.~Reinhardt,
\newblock Phys.\ Rev.\ A {\bf 72}, 63624 (2005).
\bibitem{Cederbaum1}
L.~S. Cederbaum, O.~E. Alon, and A.~I. Streltsov,
\newblock Phys.\ Rev.\ A {\bf 73}, 043609 (2006).
\bibitem{Cederbaum2}
A.~Streltsov, O.~Alon, and L.~Cederbaum,
\newblock Phys. Rev. A {\bf 73}, 063626 (2006).
\bibitem{Landau&Binder}
D.~Landau and K.~Binder,
\newblock {\em A Guide to Monte-Carlo Simulations in Statistical Physics},
\newblock Cambridge University Press, Cambridge, 2001.
\bibitem{holzmann:99}
M.~Holzmann, W.~Krauth, and M.~Naraschewski,
\newblock Phys. Rev. A {\bf 59}, 2956 (1999).
\bibitem{nilsen:05}
J.~K. Nilsen, J.~Mur-Petit, M.~Guilleumas, M.~Hjorth-Jensen, and A.~Polls,
\newblock Phys.\ Rev.\ A {\bf 71}, 053610 (2005).
\bibitem{blume:01}
D.~Blume and C.~H. Greene,
\newblock Phys. Rev. A {\bf 63}, 63061 (2001).
\bibitem{MSTTV}
A.~Minguzzi, S.~Succi, F.~Toschi, M.~Tosi, and P.~Vignolo,
\newblock Phys. Rep. {\bf 395}, 223 (2004).
\bibitem{Nunes}
G.~S. Nunes,
\newblock J.\ Phys.\ B {\bf 32}, 4293 (1999).
\bibitem{banerjee:01}
A.~Banerjee and M.~P. Singh,
\newblock Phys. Rev. A {\bf 64}, 063604 (2001).
\bibitem{GBSS}
T.~Gasenzer, J.~Berges, M.~Schmidt, and M.~Seco,
\newblock Phys.\ Rev.\ A {\bf 72}, 63604 (2005).
\bibitem{normalmode}
M.~Dunn, D.~Watson, and J.~Loeser,
\newblock Ann.\ Phys.\ (NY) {\bf 321}, 1939 (2006).
\bibitem{density0}
W.~B. Laing, M.~Dunn, and D.~K. Watson,
\newblock Phys.\ Rev.\ A {\bf 74}, 063605 (2006).
\bibitem{gross:04}
J.~L. Gross and J.~Yellen, editors,
\newblock {\em Handbook of Graph Theory},
\newblock CRC Press, Boca Raton, 2004.
\bibitem{multigraph}
Strictly speaking, this is a ``loop multigraph''. The definition of a graph
  does not allow for multiple edges between a pair of vertices nor a ``loop''
  edge with common endpoints.
\bibitem{JMPEPAPS}
W.B. Laing, M. Dunn, and D.K. Watson,
  http://nhn.ou.edu/$\sim$watson/papers/jmp08/EPAPSlaing08.pdf; Submitted to
  the EPAPS electronic depository, see http://www.aip.org/pubservs/epaps.html.
\bibitem{kellecapstone}
D.~Kelle,
\newblock Binary invariants are a basis,
\newblock Unpublished, 2008.
\bibitem{GFpaper}
B.~McKinney, M.~Dunn, D.~Watson, and J.~Loeser,
\newblock Ann. Phys. (NY) {\bf 310}, 56 (2003).
\bibitem{energy}
B.~McKinney, M.~Dunn, and D.~Watson,
\newblock Phys. Rev. A {\bf 69}, 053611 (2004).
\bibitem{mathematica}
Wolfram \protect\mbox{Research}, \textsc{Mathematica} edition: Version 6.0,
  2007.
\end{thebibliography}
 \end{document}